# Atmospheric neutrinos

F. Ronga [a]

[a] INFN Laboratori Nazionali di Frascati,
P.O. Box 13 I-00044 Frascati Italy

The results of experiments on atmospheric neutrinos are summarized, with the important exception of Superkamiokande. The main emphasis is given to the Soudan-2 and MACRO experiments. Both experiments observe atmospheric neutrino anomalies in agreement with $\nu_\mu \to \nu_\tau$ oscillations with maximum mixing. The $\nu_\mu \to \nu_s$ oscillation is disfavored by the MACRO experiment.

## 1. INTRODUCTION

Atmospheric neutrinos are produced in the cascade originated in the atmosphere by a primary cosmic ray. Underground detection of atmospheric neutrino-induced events was pioneered by the Kolar Gold Field KGF[1] experiment in India and the CWI2[2] experiment in a mine in South Africa. The field gained new interest when the large underground detectors for proton decay experiments were put in operation. At the beginning atmospheric neutrinos were studied mainly as possible sources of backgrounds for proton decay searches. But very soon, the water Cherenkov experiments, IMB in the United States and Kamiokande in Japan, discovered that the ratio between events with a muon and those with an electron was lower than expected.

The first historical observation of the anomaly was in 1986 in the IMB paper "Calculation of Atmospheric Neutrino-Induced Backgrounds in a Nucleon Decay Search"[3]. It was observed in this paper that "The simulation predicts that $34\% \pm 1\%$ of the events should have an identified muon decay while our data has $24\% \pm 3\%$". The importance of this discrepancy as possible signature for neutrino oscillations in the path length between the production point and the detector (in the range 10 -13000 km) was not fully recognized at the beginning. In 1988[5] there was the first paper by the Kamiokande collaboration dedicated to this anomaly followed by two papers from the IMB collaboration[4].

However, this anomaly was not confirmed by the proton decay iron fine-grained experiments NUSEX[6] (in the Mont Blanc tunnel between France and Italy) and Frejus[7] (in another tunnel under the Alps) and there was the suggestion that the anomaly was due to the differences in the neutrino cross sections in water and in iron not taken into account in the Fermi gas model used in the original calculations. A calculation by Engel[8] showed that this effect should be negligible for the energies of interest. Later, the results from another fine-grained iron detector Soudan-2[9] have shown that probably there was a statistical fluctuation in the NUSEX and Frejus data.

In 1994 another anomaly was observed by the Kamiokande experiments[10]: the distortion of the angular distribution of the events with a single muon in the so-called internally produced Multi-GeV data sample with a reduction of the flux of the vertical up-going events.

There were several attempts to look for possible angular distortion in other categories of events, for example in the neutrino externally produced upward-going muons. Results were produced at that time by the IMB experiment[11], the Baksan[12] experiment in URSS and the Kamiokande experiment itself[13]. The results were inconclusive or in contradiction with the neutrino oscillation hypothesis, particularly for what concerns the analysis of the stopping muon/ through-going muon ratio in the IMB experiment[11].

The MACRO tracking experiment in the Gran Sasso laboratory began the operation for neutrino physics in 1989 with a small fraction of the final detector. The first results of MACRO[14] in 1995 showed that there was a deficit of events particularly in the vertical direction. However the statis-tics was not enough at that time to discriminate unambiguously between the oscillation and the no-oscillation hypothesis.

Another big step forward in this field was due to the Superkamiokande experiment. In 1998 at the Takayama Neutrino conference there was the announcement of the observation of neutrino oscillation ($\nu_\mu$ disappearance ) from the Superkamiokande experiment. It is notable that, at the same conference, the two other running experiments Soudan-2 and MACRO have presented results in strong support of the same $\nu_\mu$ oscillations pattern observed by SuperKamiokande[15].

## 2. NEUTRINO OSCILLATIONS AND MATTER EFFECT

Neutrino oscillations[16] were suggested by B. Pontecorvo in 1957 after the discovery of the $K^0 \leftrightarrow \overline{K^0}$ transitions.

If neutrinos have masses, then a neutrino of definite flavor, $\nu_\ell$, is not necessarily a mass eigenstate. In analogy to the quark sector the $\nu_\ell$ could be a coherent superposition of mass eigenstates.

The fact that a neutrino of definite flavor is a superposition of several mass eigenstates, whose differing masses $M_m$ cause them to propagate differently, leads to neutrino oscillations : the transformation in vacuum of a neutrino of one flavor into one of a different flavor as the neutrino moves through empty space. The amplitude for the transformation $\nu_\ell \to \nu_{\ell'}$ is given by:

$$A(\nu_\ell \to \nu_{\ell'}) = \sum_m U_{\ell m} e^{-i \frac{M_m^2}{2} \frac{L}{E}} U^*_{\ell' m} \qquad (1)$$

where $U$ is a 3×3 unitary matrix in the hypothesis of the 3 standard neutrino flavors ($\nu_\mu, \nu_e, \nu_\tau$). In the hypothesis of a sterile neutrino[17] $U$ is a 4×4 unitary matrix.

The probability $P(\nu_\ell \to \nu_{\ell'})$ for a neutrino of flavor $\ell$ to oscillate in vacuum into one of flavor $\ell'$ is then just the square of this amplitude. For two neutrino oscillations and in vacuum:

$$P(\nu_\ell \to \nu_{\ell' \neq \ell}) = \sin^2 2\theta \, \sin^2 \left[ 1.27 \, \delta M^2 \frac{L}{E} \right] \qquad (2)$$

$$\delta M^2 (\text{eV}^2), L(\text{km}), E(\text{GeV})$$

This simple relation should be modified when a neutrino propagates through matter and when there is a difference in the interactions of the two neutrino flavors with matter[18].

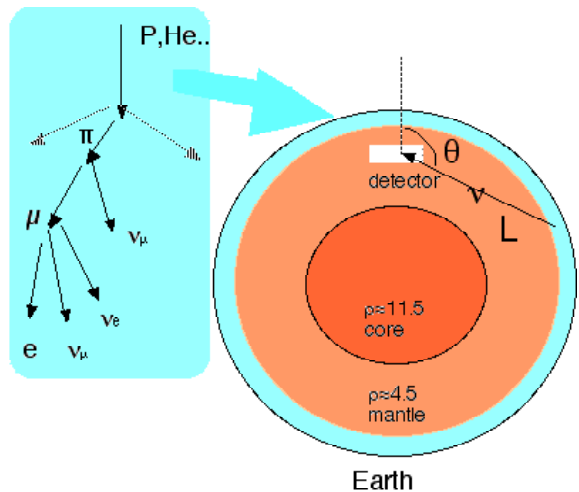

Figure 1. Sketch of the atmospheric neutrino production in the atmosphere and of the detection in an underground detetector. L is the neutrino path length and $\theta$ is the zenith angle.

The neutrino weak potential in matter is:

$$V_{\text{weak}} = \pm \frac{G_F n_B}{2\sqrt{2}} \times \begin{cases} -2Y_n + 4Y_e & \text{for } \nu_e, \\ -2Y_n & \text{for } \nu_{\mu,\tau}, \\ 0 & \text{for } \nu_s, \end{cases} \quad (3)$$

where the upper sign refers to neutrinos, the lower sign to antineutrinos, $G_F$ is the Fermi constant, $n_B$ the baryon density, $Y_n$ the neutron and $Y_e$ the electron number per baryon (both about 1/2 in normal matter). Numerically we have

$$\frac{G_F n_B}{2\sqrt{2}} = 1.9 \times 10^{-14} \text{ eV} \; \frac{\rho}{\text{g cm}^{-3}}. \qquad (4)$$

The weak potential in matter produces a phase shift that could modify the neutrino oscillation pattern if the oscillating neutrinos have different interactions with matter. The matter effect could help to discriminate between different neutrino channels. According to equation 3 the matter effect in the Earth could be important for $\nu_\mu \to \nu_e$ and for the $\nu_\mu \to \nu_s$ oscillations, while for $\nu_\mu \to \nu_\tau$ oscillations there is no matter effect. For some particular values of the oscillation parameters the matter effect could enhance the oscilla-tions originating "resonances" (MSW effect)[18]. The internal structure of the Earth could have an important role in the resonance pattern [19]. However, for maximum mixing, the only possible effect is the reduction of the amplitude of oscillations.

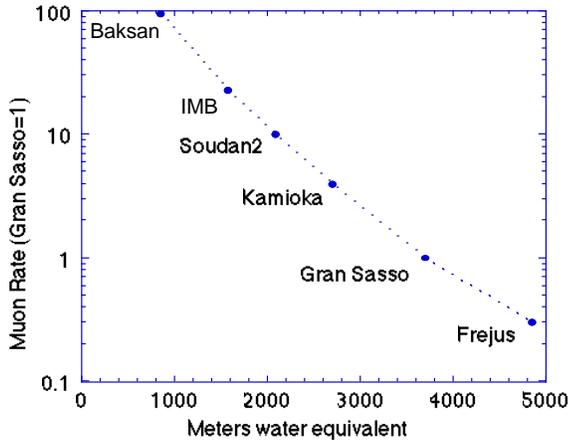

Figure 2. Rate of cosmic rays as function of the depth relative to the Gran Sasso Laboratory.

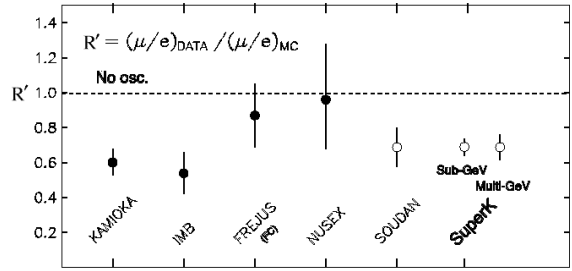

Figure 3. Measurements of the atmospheric neutrino flavor ratio[23].

## 3. ATMOSPHERIC NEUTRINOS

In the hadronic cascade produced from the primary cosmic ray we have the production of neutrinos with the following basic scheme :

$P+N \longrightarrow \pi + K..$
$\pi/k \longrightarrow \mu^+(\mu^-) + \nu_\mu \ (\overline{\nu}_\mu)$
$\mu^+(\mu^-) \longrightarrow e^+(e^-) + \nu_e(\overline{\nu_e}) + \overline{\nu_\mu}(\nu_\mu)$

From these decay channels one expects at low energies about twice muon neutrinos respect to electron neutrinos. This result doesn't change very much with a detailed calculation. The calculation of the absolute neutrino fluxes is a more complicated matter, with several sources of uncertainty[20] due to the complicated shower development in the atmosphere and to the large uncertainties in the cosmic ray spectrum.

There are two basic topologies of neutrino induced events in a detector: internally produced events and externally produced events. The internally produced events have neutrino interaction vertices inside the detector. In this case all the secondaries can be in principle observed. The range of neutrino energies involved goes from a fraction of GeV up to 10 GeV or more. Both electron neutrinos and muon-neutrinos can be detected. The externally produced events have neutrino interaction vertex in the rock below the detector. Typical neutrino energies involved are of the order of 100 GeV. Only muon neutrinos can be detected. Figure 1 shows the basic geometrical factors of the neutrino production and detection in an underground detector.

The neutrino events could have background connected with the production of hadrons by photoproduction due to the down-going muons. This background has been measured by the MACRO experiment[21]. The photoproduced neutrons can simulate internal events and the pions can simulate stopping or through-going muons. The rate of this background depends on the rate of the down-going muons and therefore on the depth. As it his shown in Figure 2 this effect could be important for detectors of shallow depth and it could be one of the reason for some past results in contradiction with the current oscillation scenario.

## 4. THE SOUDAN-2 EXPERIMENT

The results of the Soudan-2 experiment are discussed in detail in another talk at this conference[22]. Here I want to stress the importance of this experiment for the Sub-GeV events (events having energies of the order of 1 GeV or less) where a possible contradiction between the iron sampling calorimeters and the water Cherenkov detector was suggested in the past. Figure 3 shows the current experimental situation together with the new results of Soudan-2[23][24] with 4.6 Ktons/year of data.

Recently the Soudan-2 group has been able to study the L/E distribution for a sample of events selected to have an high energy resolution. The distributions are shown in Figure 4. After background subtraction they have 90.5 $\nu_\mu$ CC events and 116.4 $\nu_e$ CC (153.6 predicted) events.

Due to the nuclear effects and to the limited statistics it is not possible to see the sinusoidal pattern of equation (2) with the first minimum at $Log(\frac{L}{E}) = 2.5$ predicted in the case of oscillations with $\delta m^2 \sim 3 \times 10^{-3} eV^2$. One of the main goal of the next generation atmospheric neutrino experiments is the measurement of this pattern that could provide a precise measurement of the oscillation parameters and a way to discriminate alternative hypothesis with neutrino decays[25].

However, from the study of the $\chi^2$ of the L/E distribution as function of the oscillation parameters the Soudan-2 group has been able to set a 90% confidence region for the oscillation parameters, reported together with the other experiments in figure 12. The best $\chi^2$ is for $\delta m^2 = 0.8 \times 10^{-2} eV^2$ and $sin^2 2\theta = 0.95$.

Soudan-2 measures a flux of $\nu_e$ neutrinos smaller than the one expected (with an old version of the Bartol flux), while SuperKamiokande has agreement between predictions and data. This disagreement could be due or to a statistical fluctuation or to some physical effect due to the different geomagnetic cuts or to differences in the neutrino samples (Soudan-2 accepts events with a smaller energy and accepts multiprong events).

## 5. NEUTRINO DETECTION IN THE MACRO EXPERIMENT

The MACRO detector is described elsewhere[14][30]. Active elements are streamer tube chambers used for tracking and liquid scintillator counters used for the time measurement.

Figure 5 shows a schematic plot of the three different topologies of neutrino events analyzed up to now: *Up Through* events, *Internal Up* events and *Internal Down* together with *Up Stop* events. The requirement of a reconstructed track selects events having a muon.

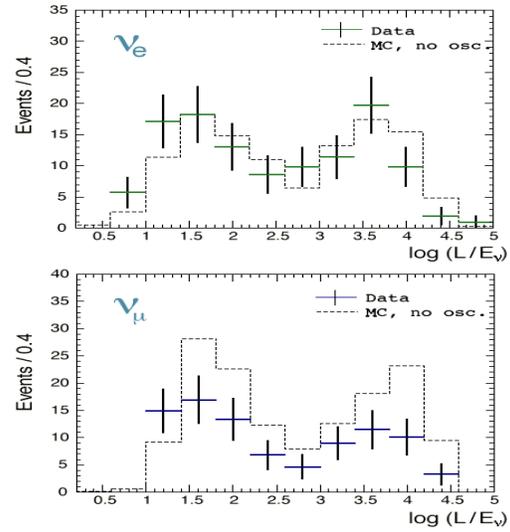

Figure 4. $\frac{L}{E}$ distribution for the Soudan-2 experiment[23][22]. Top: $\nu_e$ with data normalized to the prediction. Bottom : $\nu_\mu$ with normalization taken from $\nu_e$.

The *Up Through* tracks come from $\nu_\mu$ interactions in the rock below MACRO. The muon crosses the whole detector ($E_\mu > 1$ GeV). The time information provided by scintillator counters permits to know the flight direction (time-of-flight method). The typical neutrino energy for this kind of events is of the order of 100 GeV. The data have been collected in three periods, with different detector configurations starting in 1989 with a small fraction of the apparatus.

The *Internal Up* events come from $\nu$ interactions inside the apparatus. Since two scintillator layers are intercepted, the time-of-flight method is applied to identify the upward going events. The typical neutrino energy for this kind of events is around 4 GeV. If the atmospheric neutrino anomalies are the result of $\nu_\mu$ oscillations with maximum mixing and $\Delta m^2$ between $10^{-3}$ eV$^2$ and $10^{-2}$ eV$^2$ it is expected a reduction in the flux of this kind of events of about a factor of two, without any distortion in the shape of the angular distribution. Only the data collected with the full MACRO (live-time around 4.1 years) have been used in this analysis.

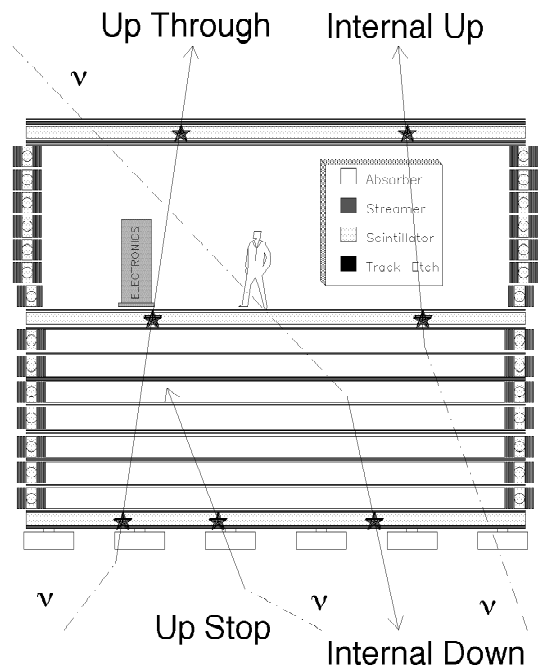

Figure 5. Sketch of different event topologies induced by neutrino interactions in or around MACRO (see text). In the figure, the stars represent the scintillator hits. The time of flight of the particle can be measured only for the *Internal Up* and *Up Through* events.

The *Up Stop* and the *Internal Down* events are due to external interactions with upward-going tracks stopping in the detector (*Up Stop*) and to neutrino induced downgoing tracks with vertex in the lower part of MACRO (*Internal Down*). These events are identified by means of topological criteria. The lack of time information prevents to distinguish the two sub samples. The data set used for this analysis is the same used for the *Internal Up* search. An almost equal number of *Up Stop* and *Internal Down* is expected if neutrinos do not oscillate. The average neutrino energy for this kind of events is around 4 GeV. In case of oscillations it is not expected a reduction in the flux of the *Internal Down* events (having path lengths of the order of 20 km), while it is expected a reduction in the number *Up Stop* events similar to the one expected for the *Internal Up*.

## 6. UPWARD THROUGH-GOING MUONS

The direction that muons travel through MACRO is determined by the time-of-flight between two different layers of scintillator counters. The measured muon velocity is calculated with the convention that muons going down through the detector are expected to have $1/\beta$ near $+1$ while muons going up through the detector are expected to have $1/\beta$ near -1.

Several cuts are imposed to remove backgrounds caused by radioactivity or showering events which may result in bad time reconstruction. The most important cut requires that the position of a muon hit in each scintillator as determined from the timing within the scintillator counter agrees within $\pm 70$ cm with the position indicated by the streamer tube track.

In order to reduce the background due to the photoproduced pions each upgoing muon must cross at least 200 g/cm$^2$ of material in the bottom half of the detector. Finally, a large number of nearly horizontal ($\cos \theta > -0.1$), but upgoing muons have been observed coming from azimuth angles corresponding to a direction containing a cliff in the mountain where the overburden is insufficient to remove nearly horizontal, downgoing muons which have scattered in the mountain and appear as upgoing. This region is excluded from both the observation and Monte-Carlo calculation of the upgoing events.

There are 561 events in the range $-1.25 < 1/\beta < -0.75$ defined as upgoing muons for this data set. These data are combined with the previously published data [14] for a total of 642 upgoing events. Based on the events outside the upgoing muon peak, it is estimated there are $12.5 \pm 6$ background events in the total data set. In addition to these events, there are $10.5 \pm 4$ events which result from upgoing charged particles produced by downgoing muons in the rock near MACRO. Finally, it is estimated that $12 \pm 4$ events are the result of interactions of neutrinos in the very bottom layer of MACRO scintillators.

In the upgoing muon simulation it is used the neutrino flux computed by the Bartol group[26]. The cross-sections for the neutrino interactions

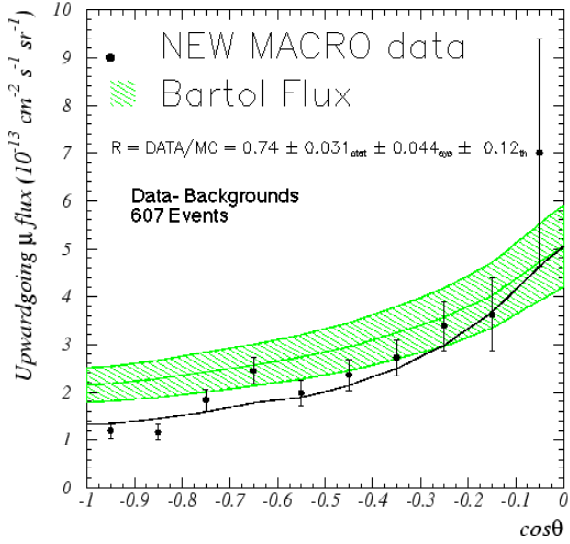
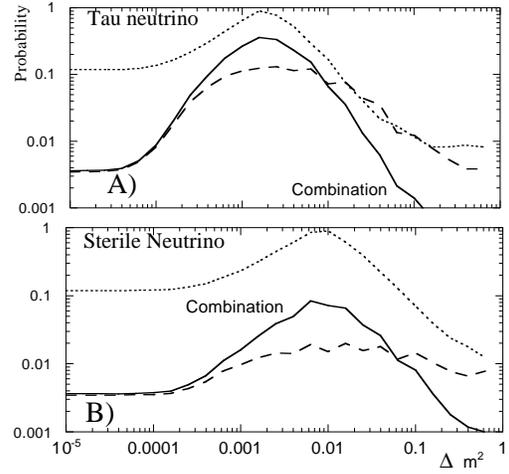

Figure 6. Zenith distribution of flux of upgoing muons with energy greater than 1 GeV for data and Monte Carlo for the combined MACRO data. The shaded region shows the expectation for no oscillations with the 17% uncertainty in the expectation. The solid line shows the prediction for an oscillated flux with $\sin^2 2\theta = 1$ and $\Delta m^2 = 0.0025$ eV$^2$.

Figure 7. A: Probabilities for maximum mixing and oscillations $\nu_\mu \to \nu_\tau$. B: oscillations $\nu_\mu \to \nu_s$. The 3 lines corresponds to the probability from the total number of events (dotted line), the probability from the chi-square of the angular distribution with data and prediction normalized (dashed line) and to the combination of the two independent probabilities (continous line)

have been calculated using the GRV94[27] parton distributions set which varies by +1% respect to the Morfin and Tung parton distribution used in the past. The systematic error on the upgoing muon flux due to uncertainties in the cross section including low-energy effects[28] is 9%. The propagation of muons to the detector has been done using the energy loss calculation of reference[29] for standard rock. The total systematic uncertainty on the expected flux of muons adding the errors from neutrino flux, cross-section and muon propagation in quadrature is ±17%. This theoretical error in the prediction is mainly a scale error that doesn't change the shape of the angular distribution. The number of events expected integrated over all zenith angles is 824.6, giving a ratio of the observed number of events to the expectation of 0.74 ±0.031(stat) ±0.044(systematic) ±0.12(theoretical).

Figure 6 shows the zenith angle distribution of the measured flux of upgoing muons with energy greater than 1 GeV for all MACRO data compared to the Monte Carlo expectation for no oscillations and with a $\nu_\mu \to \nu_\tau$ oscillated flux with $\sin^2 2\theta = 1$ and $\Delta m^2 = 0.0025$ eV$^2$.

The shape of the angular distribution has been tested with the hypothesis of no oscillations normalizing data and predictions. The $\chi^2$ is 22.9, for 8 degrees of freedom (probability of 0.35% for a shape at least this different from the expectation). Also $\nu_\mu \to \nu_\tau$ oscillations are considered. The best $\chi^2$ in the physical region of the oscillations parameters is 12.5 for $\Delta m^2$ around $0.0025 eV^2$ and maximum mixing (the best $\chi^2$ is 10.6, outside the physical region).

To test the oscillation hypothesis, the independent probabilities for obtaining the number of events observed and the angular distribution for various oscillation parameters are calculated. They are reported for $\sin^2 2\theta = 1$ in Figure 7 A) for oscillations $\nu_\mu \to \nu_\tau$. It is notable that the value of $\Delta m^2$, suggested from the shape of the angular distribution is similar to the value necessary in order to obtain the observed reduction in the total number of events in the hypothesis of maximum mixing. Figure 7 B) shows the same quantities for sterile neutrinos oscillations taking into account matter effects.

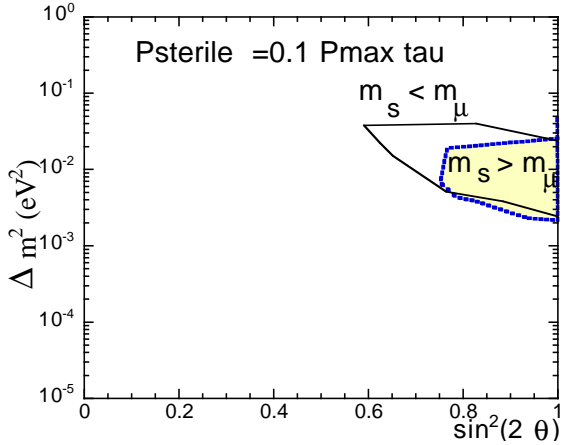

Figure 8. Iso-probability plot for sterile neutrino showing the contours corresponding to a 3.6% probability (10% of the maximum probability for $\nu_\tau$). The two regions are for positive and negative values of $\Delta m^2$

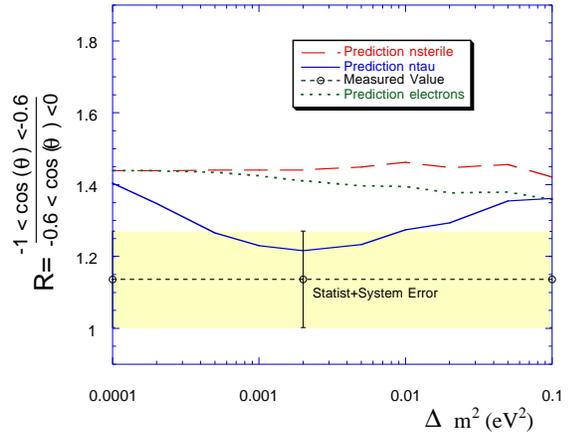

Figure 9. The ratio obtained binning the data in two bins and the comparison with the $\nu_s,\nu_e$ and $\nu_\tau$ oscillations with maximum mixing

The maximum of the probability is 36.6% for oscillations $\nu_\mu \to \nu_\tau$. The probability for no oscillation is 0.36%.

The maximum probability for the sterile neutrino is 8.6% in a region of $\Delta m^2$ around $10^{-2} eV^2$. The probabilities for sterile neutrinos are comparable to the one for $\tau$ neutrinos only in the small regions shown in Figure 8.

Another way to try to discriminate between the oscillation of $\nu_\mu$ in $\nu_s$ and the one in $\nu_\tau$ is to study the angular distribution in two bins, computing the ratio between the two bins as shown in Figure 9. The statistical significance is higher then in the case of data binned in 10 bins, but some features of the angular distribution could be lost using this ratio. The ratio is insensitive to most of the errors on the theoretical prediction of the $\nu$ flux and cross section[31]. From this plot the $\nu_s$ hypothesis is disfavored at $2\sigma$ level.

It is interesting to check if there is agreement between the data measured by different experiments. In case of oscillations it is important to take into account correctly the energy threshold of the different experiments: Superkamiokande has an average energy threshold of the order of 7 GeV while for MACRO it is 1 GeV. The comparison between Kamiokande[32], Superkamiokande[33] and MACRO shown in Figure 10 is done by comparing the ratio between the events measured and the events expected in case of oscillation (as computed by each experiment). There is a remarkable agreement between the three experiments even if there is still some possible discrepancy in the region around the vertical.

## 7. THE MACRO LOW ENERGY EVENTS

The analysis of the *Internal Up* events is similar to the analysis of the *Up Through*. The main difference is due to the requirement that the interaction vertex should be inside the apparatus. About 87% of events are estimated to be $\nu_\mu$ CC interactions. The uncertainty due to the acceptance and analysis cuts is 10%. After the background subtraction (5 events) 116 events are classified as *Internal Up* events

The *Internal Down* and the *Up Stop* events

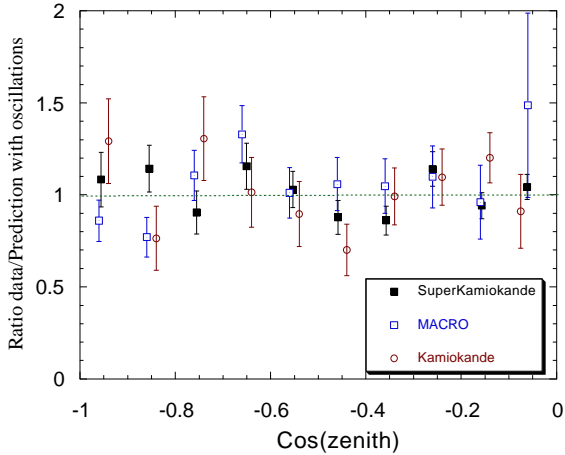

Figure 10. Comparison of the upward-going through-going muon data of Kamiokande[32], Superkamiokande[23], MACRO. The prediction in case of oscillation (as computed from each experiment for the best fit point) are for different value of $\delta m^2$ and (for Kamiokande and Superkamiokande) for slightly different values of the normalization.

are identified via topological constraints. The main requirement is the presence of a reconstructed track crossing the bottom scintillator layer. The tracking algorithm for this search requires at least 3 streamer hits (corresponding roughly to 100 $gr\ cm^{-2}$). All the track hits must be at least 1 m from the detector's edges. The criteria used to verify that the event vertex (or stopping point) is inside the detector are similar to those used for the *Internal Up* search. To reject ambiguous and/or wrongly tracked events which survived automated analysis cuts, real and simulated events were randomly merged and directly scanned with the MACRO Event Display. About 90% of the events are estimated to be $\nu_\mu$ CC interactions. The main background for this search are the low energy particles produced by donwn-going muons [21]. After background subtraction ($7 \pm 2$ events) 193 events are classified as *Internal Down* and *Up Stop* events. The Montecarlo simulation for the low energy events uses the Bartol neutrino flux [26] and the neutrino low energy cross sections reported in [28]. The simulation is performed in a large volume of rock (170 kton) around the MACRO detector (5.3 kton). The uncertainty on the expected muons flux is about 25%. The number of expected events was also evaluated using the "NEUGEN" neutrino event generator [34] (developed by the Soudan and MINOS collaborations) as input to our full Monte Carlo simulation. The NEUGEN generator predicts $\sim 6\%(5\%)$ fewer IU (ID+UGS) events detectable in MACRO than [28], well within the estimated systematic uncertainty for neutrino cross sections ($\sim 15\%$).

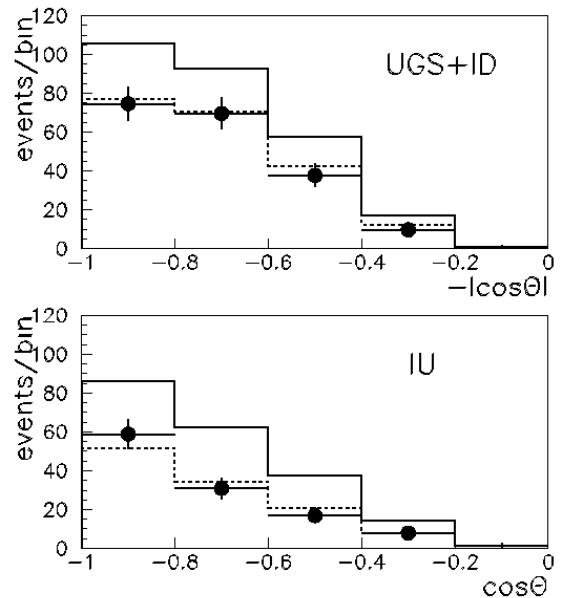

Figure 11. Zenith angle ($\theta$) distribution for $IU$ and $UGS + ID$ events. The background-corrected data points (black points with error bars) are compared with the Monte Carlo expectation assuming no oscillation (full line) and two-flavor oscillation (dashed line) using maximum mixing and $\Delta m^2 = 2.5 \times 10^{-3}\ eV^2$.

The angular distributions of data and predictions are compared in Figure 11. The low energy samples show an uniform deficit of the measured number of events over the whole angular distribution with respect to the predictions, while there is good agreement with the predictions based on neutrino oscillations.

The theoretical errors coming from the neu-trino flux and cross section uncertainties almost cancel if the ratio between the measured number of events $\frac{IU}{(ID+UGS)}$ is compared with the expected one. The partial error cancellation arises from the nearly equal energy spectra of parent neutrinos for the IU and the ID+UGS events. The experimental systematic uncertainty on the ratio is 6%. The measured ratio is $\frac{IU}{ID+UGS} = 0.60 \pm 0.07_{stat}$, while the one expected without oscillations is $0.74 \pm 0.04_{sys} \pm 0.03_{theo}$. The probability (one-sided ) to obtain a ratio so far from the expected one is 5%, near independent of the used neutrino flux and neutrino cross sections for the predictions.

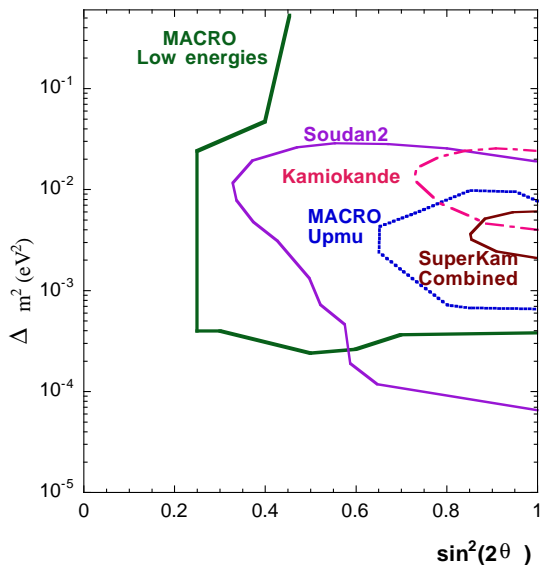

Figure 12. The 90% confidence level regions of the experiments with positive indication of oscillation for atmospheric neutrinos. The MACRO limits are computed using the Feldman-Cousins procedure[35]

The confidence level regions as function of $\Delta m^2$ and $sin^2 2\theta$ are studied using a $\chi^2$ comparison of data and Monte Carlo for the data of Fig. 11. The $\chi^2$ includes the shape of the angular distribution, the $\frac{IU}{ID+UGS}$ ratio and the overall normalization. The systematic uncertainty is 10% in each bin of the angular distributions, while it is 5% for the ratio. The result is shown in Figure 12 (MACRO low energes).

A direct comparison can be done with the Superkamiokande stopping muons [36], because in this analisys it is used the same Bartol neutrino flux used from MACRO. The ratio $\frac{measured}{expected}$ with oscillations in the best fit point of each experiment and for $cos(\theta) < -0.2$ is $0.97 \pm 0.11$ for SuperKamiokande in good agreement with $1.02 \pm 0.15$ for the MACRO $IUP$ and $0.91 \pm 0.11$ for the MACRO $IDW + STOP$ events.

## 8. CONCLUSIONS

The Soudan-2 detector is able to study atmospheric neutrino oscillations in the Sub-GeV region. MACRO is able to cover the Multi-GeV and the $\sim 100 GeV$ region. MACRO and Soudan2 results can be compared to Kamiokande and SuperKamiokande that covers all the three region. The 90% allowed confidence regions for Kamiokande,Superkamiokande, Soudan-2, MACRO for the oscillation $\nu_\mu \to \nu_\tau$ are shown in Figure 12. The statistical power of the Superkamiokande experiment is larger than the others, but it is remarkable to note that it is possible to see the same effect detected in Superkamiokande with detectors using completely different experimental techniques and in similar energy regions.

Using the matter effect it is possible to discriminate between different oscillation hypothesis. In particular the oscillation $\nu_\mu \to \nu_s$ with maximum mixing is disfavored by the MACRO experiment respect the $\nu_\mu \to \nu_\tau$ oscillation. A similar results is obtained by Superkamiokande.

Although the $\nu_\mu \to \nu_\tau$ oscillation hypothesis is the most simple that fits the current data, other more complex scenarios exist and can fit well the data. They require additional hypothesis on the existence of exotic particles, such as the sterile neutrino, or of oscillations in more than 2 families[37].

The exact determination of oscillation parameters and of the channels of the oscillations will be the main goal of the future generation experiments using atmospheric neutrinos or artificial neutrino beams.

I am greatly indebted to W. A. Mann, Maury Goodman and T. Kafka of the Soudan2 Group and all the collegues of the MACRO collaboration for the results presented in this Talk and for the very useful discussions.

## REFERENCES

1. C. V. Achar et al Phys Letters 18, 196 and 1978 (1965)
2. F Reines et al., Phys Rev Letters 15, (1965) 429
3. T. J Heines et al (IMB collaboration ) Phys Rev Lett 57, 1986 (1986)
4. D.Casper et al.(IMB collaboration) Phys Rev Lett 66 2561 (1991) , R. Becker-Szendy et al. Phys Rev D 46 3720 (1992)
5. K. S. Hirata et al (Kamiokande collaboration) Phys Lett B 205 (1988) 416 and Phys Letters B280 (1992) 146
6. M Aglietta et al (NUSEX collaboration) Europhys Letters 8 (1989) 611
7. Ch Berger et al. (Frejus collaboration) Phys Lett B227, 489 (1989) and Phys Letters B245 (1990)305.
8. J. Engel, E. Kolbe, K. Lagangke and P. Vogel Phys Rev D 48 (1993) 3048.
9. W.W. Allison *et al.*, (Soudan-2 Collaboration) Phys. Lett. B391, 491 (1997) hep-ex/9611007., W.W. Allison *et al.* Nucl. Phys. Proc. Suppl. 35, 427 (1994).
10. Y. Fukuda et al. (Kamiokande collaboration) Phys Lett B 235 237 (1994)
11. R Becker-Szendy et al Phys Rev. Lett. 69 1010 (1992)
12. S. P Mikheiev (Baksan collaboration) proceedings of the 24th Cosmic ray Conference - Rome 1 722 (1995), Boliev et al. (Baksan collaboration) Nucl. Phys. Proc. Suppl. 70 (1999) 371.
13. Y. Oyama *et al.* (KAMIOKANDE-II Collaboration), Phys. Rev. D39, 1481 (1989). W. Frati, T.K. Gaisser, A.K. Mann and T. Stanev, Phys. Rev. D48, 1140 (1993).
14. S. Ahlen et al. (MACRO collaboration) Phys. Lett. B 357 (1995) 481, D. Michael (MACRO collaboration) Nucl. Phys. Proc. Suppl. 35 (1994) 235, F Ronga (MACRO collaboration) Helsinki 1996 Neutrino Conference World Scientific.
15. T. Kajita (Super-Kamiokande Collaboration), Nucl. Phys. Proc. Suppl. 77, 123 (1999), E. Peterson (Soudan-2 Collaboration), Nucl. Phys. Proc. Suppl. 77, 111 (1999), F. Ronga *et al.* (MACRO Collaboration), Nucl.Phys.Proc.Suppl. 77 117,(1999).
16. B. Pontecorvo J.Exptl. Theoret. Phys. 33, 549 (1957), Z. Maki, M. Nakagava and S. Sakata, Prog. Theor. Phys. 28, 870 (1962), for an historical review see S. M. Bilenki hep-ph/ 9908335 (1999).
17. E. Akhmedov, P. Lipari, and M. Lusignoli, Phys. Lett. B 300, 128 (1993), F. Vissani and A.Y. Smirnov, Phys. Lett. B432, 376 (1998) Q.Y. Liu and A.Y. Smirnov, Nucl. Phys. B524, 505 (1998) P. Lipari and M. Lusignoli, Phys. Rev. D58, 073005 (1998)
18. Wolfenstein L. Phys. Rev. D17:2369 (1978), Phys. Rev D20:2634 (1979), Mikheyev SP, Smirnov AYu. Sov. J. Nucl. Phys. 42:913 (1985), Sov. Phys. -JETP64:4 (1986), Nuovo Cimento 9C:17 (1986)
19. P.I. Krastev and A.Y. Smirnov, Phys. Lett. B226, 341 (1989). Q.Y. Liu, S.P. Mikheyev and A.Y. Smirnov, Phys. Lett. B440, 319 (1998) M.V. Chizhov and S.T. Petcov, Phys. Rev. Lett. 83, 1096 (1999)
20. T. Gaisser: talk at this conference.
21. Ambrosio, M., et al. (MACRO collabor.) Astroparticle Physics 9 (1998) 105.
22. Kafka, T (Soudan-2 Collab.):talk at this conference.
23. Mann W.A.: Talk at the 1999 Lepton Photon conference hep-ex/9912007
24. W.W. Allison *et al.* (Soudan-2 Collab.), Phys. Lett. B449, 137 (1999)
25. K. Hoepfner: Talk at this conference
26. Agrawal et al. Phys Rev D53 1314 (1996)
27. Glück M., Reya E. and Stratmann M.1995, Z. Phys. C67, 433
28. Lipari P. Lusignoli M. and Sartogo F. 1995, Phys. Rev. Lett. 74 4384
29. Lohmann H. Kopp R.,Voss R. 1985, CERN-EP/85-03
30. M. Ambrosio *et al.* (MACRO Collab.), Phys. Lett. B434, 451 (1998) hep-ex/9807005.
31. P. Lipari and M. Lusignoli, Phys. Rev. D57, 3842 (1998)
32. Hatakeyama S. et al. (Kamiokande collaboration) 1998, Phys Rev Lett 81 2016
33. M. Nakahata (SuperKamiokande Collab.) : talk at this conference
34. Private communication from the MINOS collaboration.
35. G.Feldman and R.Cousins, Phys. Rev. D57

(1998)3873.
36. Y. Fukuda *et al.* (SuperKamiokande Collaboration), Phys. Lett. B467, 185 (1999)
37. G.L. Fogli : talk at this conference, C. Giunti, Phys. Lett. B467, 83 (1999)